\documentclass[sigconf]{acmart}

\AtBeginDocument{%
  }

\copyrightyear{2025}
\acmYear{2025}
\setcopyright{cc}
\acmConference[HASP 2025]{Hardware and Architectural Support for Security and Privacy 2025}{October 19, 2025}{Seoul, Republic of Korea}
\acmBooktitle{Hardware and Architectural Support for Security and Privacy 2025 (HASP 2025), October 19, 2025, Seoul, Republic of Korea}
\acmDOI{10.1145/3768725.3768726}
\acmISBN{979-8-4007-2198-4/25/10}

\usepackage{xcolor}
\usepackage{algorithm}
\usepackage{acmart-taps}
\usepackage{tikz}
\usepackage{listings}
\usepackage{textcomp}
\usepackage{pifont}
\usepackage{multirow}
\usepackage{colortbl}
\usepackage{mydefs}
\usepackage{soul}
\usepackage[font=footnotesize,labelfont=bf]{caption}

\newcommand{\cmark}{\ding{51}}%
\newcommand{\xmark}{\ding{55}}%

\begin{document}

\title{CRAFT: \underline{C}haract\-erizing and \underline{R}oot-\-C\underline{a}using \underline{F}ault Injection \underline{T}hreats at Pre-Silicon}

\author{Arsalan Ali Malik}
\affiliation{%
  \institution{North Carolina State University}
    \city{Raleigh}
  \state{North Carolina}
  \country{USA}
}
\email{aamalik3@ncsu.edu}
\author{Harshvadan Mihir}
\affiliation{%
  \institution{North Carolina State University}
  \city{Raleigh}
  \state{North Carolina}
  \country{USA}
}
\email{hmihir@ncsu.edu}

\author{Aydin Aysu}
\affiliation{%
  \institution{North Carolina State University}
  \city{Raleigh}
  \state{North Carolina}
  \country{USA}
}
\email{aaysu@ncsu.edu}

\renewcommand{\shortauthors}{Arsalan Ali Malik et al.}
\begin{abstract}
Fault injection attacks (FIA) pose significant security threats to embedded systems as they exploit weaknesses across multiple layers, including system software, instruction set architecture (ISA), microarchitecture, and physical hardware. Early detection and understanding of how physical faults propagate to system-level behavior are essential to safeguarding cyberinfrastructure.

This work introduces CRAFT, a framework that combines pre-silicon analysis with post-silicon validation to systematically uncover and analyze fault injection vulnerabilities. Our study, conducted on a RISC-V soft-core processor (cv$32$e$40$x) reveals two novel vulnerabilities. First, we demonstrate a method to induce instruction skips by glitching the clock (single-glitch attack), which prevents critical values from being loaded from memory, thus disrupting program execution. Second, we show a technique that converts a fetched legal instruction into an illegal one mid-execution, diverting control flow in a manner exploitable by attackers. Notably, we identified a specific timing window in which the processor fails to detect these illegal control-flow diversions, allowing silent, undetected corruption of the program state. 

By simulating $9248$ FIA scenarios at pre-silicon and conducting root-cause analysis of the RISC-V pipeline, we trace the faults to a previously unreported vulnerability in a pipeline register shared between the instruction fetch and decode stages. Our approach reduced the search space for post-silicon experiments by $97.31$\%, showing pre-silicon advantages for post-silicon testing. Finally, we validate our identified exploit cases on real hardware (FPGA). 
\end{abstract}




\maketitle

\section{Introduction}
\vspace{-.25em}

Fault injection attacks (FIAs) deliberately introduce faults into a system to alter its behavior, exploiting, \textit{e.g.}, 
timing characteristics of designs to induce malfunctions. An attacker may have various goals, such as (a) bypassing security mechanisms to gain unauthorized access to sensitive resources~\cite{murdock2020plundervolt,claudepierre2021traitor}; 
(b) triggering errors during cryptographic computations to reveal encryption key~\cite{yuce2018fault,malik2024enabling}; (c) disrupting critical system functionality, causing failures in IoT applications~\cite{gangolli2022systematic,malik2020isolation,THEMIS}; (d) injecting faults in embedded firmware to execute unauthorized instructions~\cite{mutlu2019rowhammer}, and (e)  introducing (data) errors to undermine the reliability~\cite{chamelot2022sci,gongye2024one}. These attacks can target different layers of the system stack—from software down to hardware—and exploit the gap between how systems are designed for normal conditions vs. how they operate under physical stress.
\\\indent

\vspace{-1.25em}
Designers previously relied on post-silicon analysis to apply FIA countermeasures. However, rising attack sophistication and hardware complexity now necessitate a multi-stage approach across the design process. The interconnected nature of modern systems demands comprehensive protection throughout the development lifecycle, from initial design to final implementation. Therefore, to successfully defend against FIAs, vulnerabilities must be analyzed and caught at multiple stages of the design process. Pre-silicon analysis (simulation or emulation using RTL/netlist) identifies weaknesses early, allowing fixes before fabrication~\cite{SoFI,kazemi2021depth,faultdetective, SYNFI}. However, without physical testing, real-chip fault appearance remains uncertain. Conversely, post-silicon testing validates exploits on real hardware but often treats the processor as a `black box,' hindering root-cause tracing within the design.


\begin{figure}[t]
    \vspace{1.5em}
	\includegraphics[width=0.48\textwidth]{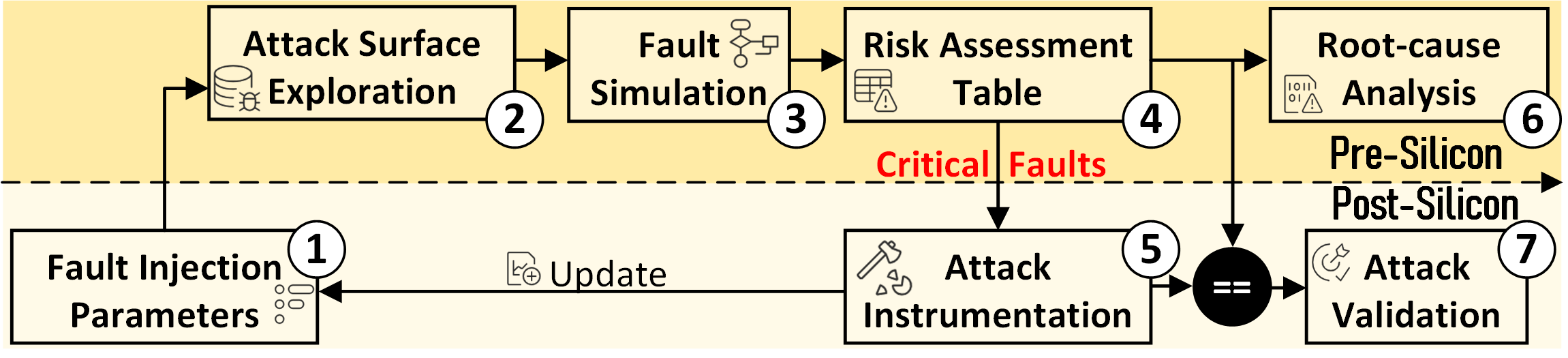}
	\vspace{-2em}
    \footnotesize
	\caption{CRAFT follows a seven-step process: \ding{172} select fault parameters, such as glitch width and glitch offset, based on post-silicon capabilities; \ding{173} explore the attack surface by analyzing the target application and identifying instructions of interest; \ding{174} inject faults in pre-silicon (simulation) to identify failing instructions; \ding{175} compile a risk assessment table (RAT) to classify and prioritize critical faults; \ding{176} use RAT to execute an attack on post-silicon; \ding{177} perform root-cause analysis on failing instructions and update fault parameters to improve accuracy for future iterations, and \ding{178} validate pre-silicon attack results against post-silicon outcomes to confirm the vulnerability.}
    \Description{CRAFT follows a seven-step process: {1} select fault parameters, such as glitch width and glitch offset, based on post-silicon capabilities; {2} explore the attack surface by analyzing the target application and identifying instructions of interest; {3} inject faults in pre-silicon (simulation) to identify failing instructions; {4} compile a risk assessment table (RAT) to classify and prioritize critical faults; {5} use RAT to execute an attack on post-silicon; {6} perform root-cause analysis on failing instructions and update fault parameters to improve accuracy for future iterations, and {7} validate pre-silicon attack results against post-silicon outcomes to confirm the vulnerability.}
\vspace{-2em}
	\label{fig: Process}
\end{figure}

While prior works aim to inject various types of faults~\cite{riviere2015high,shuvo2023comprehensive} or mitigate them by addressing their symptoms~\cite{nasahl2023scramble}, limited research has focused on identifying the root causes of FIA vulnerabilities across pre- and post-silicon stages~\cite{faultdetective,yuce2015improving}. Furthermore, a key limitation of these works is their emphasis on analyzing the effects of faults, such as instruction skips or data corruption, while providing insufficient attention to investigating the underlying causes of fault generation.
Understanding why these faults occur across multiple layers—like the instruction set architecture (ISA), microarchitecture, and physical hardware—remains critical to designing effective countermeasures and uncovering novel (undocumented) exploits. 

Figure~\ref{fig: Process} illustrates our proposed framework, CRAFT, which integrates pre-silicon fault analysis with post-silicon fault validation. CRAFT employs a seven-step process to identify, characterize, and root-cause vulnerabilities, focusing on critical pipeline control flow faults like instruction skips or unintended jumps~\cite{malik2025honest}. The proposed framework's principles, including clock glitch characterization, are fundamentally applicable across different technologies (ASIC/FPGA).
Using CRAFT, we conducted an extensive case study on the open-source RISC-V processor (cv$32$e$40$x), uncovering a new clock-glitch attack vector and its root cause. 
CRAFT systematically leverages insights from pre-silicon analysis to guide post-silicon testing, focusing on faults that affect pipeline control flow—such as instruction skips or unintended jumps. We trace how a single glitch propagates through the pipeline by conducting $9248$ fault injection simulations at the post-synthesis netlist level and reproducing a subset on real hardware. This end-to-end analysis identifies which faults occur, why they happen, and where to implement hardware defenses. 

The key contributions of this work are the following:

\begin{itemize}
    \item \textbf{RISC-V instruction vulnerability characterization.}
    We inject faults into eight representative RISC-V instructions from an embedded neural network workload through each pipeline stage. We rank instruction vulnerability, with the highest percentages representing the most vulnerable instructions and stages, to produce a risk assessment table (RAT) that highlights at-risk areas.
    
    \item \textbf{Discovery of novel fault scenarios.}
    We identify four exploit cases where precise clock glitches redirect program flow or corrupt results. These include a novel way to induce instruction skips, preventing critical operations. CRAFT converts legal instructions into illegal ones on the fly, bypassing detection. We also demonstrate cases when the processor fails to raise an illegal instruction exception, allowing the attack to remain undetected.
    
    \item \textbf{Root-cause analysis and verification.} 
    We trace and root cause the silent illegal instruction vulnerability to a specific pipeline register (IF/ID stage) and its interaction with the RISC-V compressed instruction decoder. We confirm how corrupted bits cause erroneous program counter (PC) updates and missed exception flags by analyzing simulation waveforms and RTL code. To the best of our knowledge, this is the \emph{first} report of this vulnerability in the cv$32$e$40$x core.
    
    \item \textbf{Pre- to post-silicon validation of attacks.} 
    Using insights from pre-silicon analysis, we conduct and verify targeted clock-glitch experiments on the FPGA implementation of the same RISC-V core. Our approach cut hardware experiment search space by $97.31$\%, proving design-time (pre-silicon) insights' value for post-silicon testing.
\end{itemize}
Our work presents a comprehensive approach to uncovering and understanding fault injection threats. By bridging pre-silicon analysis and post-silicon validation, CRAFT provides system designers with a template for identifying hidden hardware vulnerabilities before they can be exploited, guiding the development of effective defenses. 
While our case study centers on a specific RISC-V core and a subset of instructions, the framework is generalizable to other processors and larger instruction sets.
\section{Background}
This section outlines the key concepts relevant to our work and formalizes the threat model assumed in this work.
\subsection{The Research Gap}
\vspace{-.25em}
Although a growing body of research has aimed to analyze and mitigate fault injection threats~\cite{verbauwhede11,shuvo2023comprehensive,zhang2019recent,liu2020imperceptible}, two critical limitations persist for \emph{clock glitch attacks}.
First, these works either focus on pre-silicon~\cite{Simplifi} or post-silicon~\cite{yuce2015improving,kazemi2019low, shukla2023efficient,Yuce2016} but rarely on both~\cite{kazemi2021depth,faultdetective}. Both approaches are crucial, as pre-silicon analysis simulates internal intricacies and identifies root causes, while post-silicon validation confirms that the identified vulnerabilities persist in the final product.
For example, post-silicon characterization~\cite{yuce2015improving} can identify the effects of the faults and the attack parameters needed to achieve these effects, but it cannot identify which paths in the circuit are causing these faults, which are needed to build low-overhead defenses or eliminate issues in the current design.

Commercial emulation platforms, such as Synopsys ZeBu and Cadence Palladium, excel at high-speed functional verification and general system-level emulation~\cite{Zebu,Pladdium}. However, they do not primarily support the fine-grained, precise timing fault injection characterization critical for discovering subtle hardware vulnerabilities~\cite{balasubramanian2014understanding}. Moreover, their abstraction level lacks the granularity necessary for systematically exploring specific timing windows or performing root-cause analysis of security-critical physical faults~\cite{Simplifi}.
\\\indent
The closest works to our proposal conduct both pre- and post-silicon fault characterization of RISC-V and MSP430 ISA, 
respectively~\cite{kazemi2021depth,faultdetective}.
Kazemi \textit{et al.} conducted pre-silicon analysis using a C++ cycle-accurate model of processor behavior, 
which limits them to certain standard libraries and high-level functions.
By contrast, our work identifies and tracks the traversal of faults from critical circuit elements to the application layer.
FaultDetective investigates hardware-level faults by observing their manifested effects at the software level~\cite{faultdetective}. However, it relies on a redundant microcontroller design in lock-step\footnote{In a lock-step configuration, two or more processors (or cores) execute the same instructions simultaneously and in parallel, cycle by cycle.} and requires scan registers to observe internal states, limiting its scope and applicability. 
\begin{figure}[t]
	\includegraphics[width=0.45\textwidth]{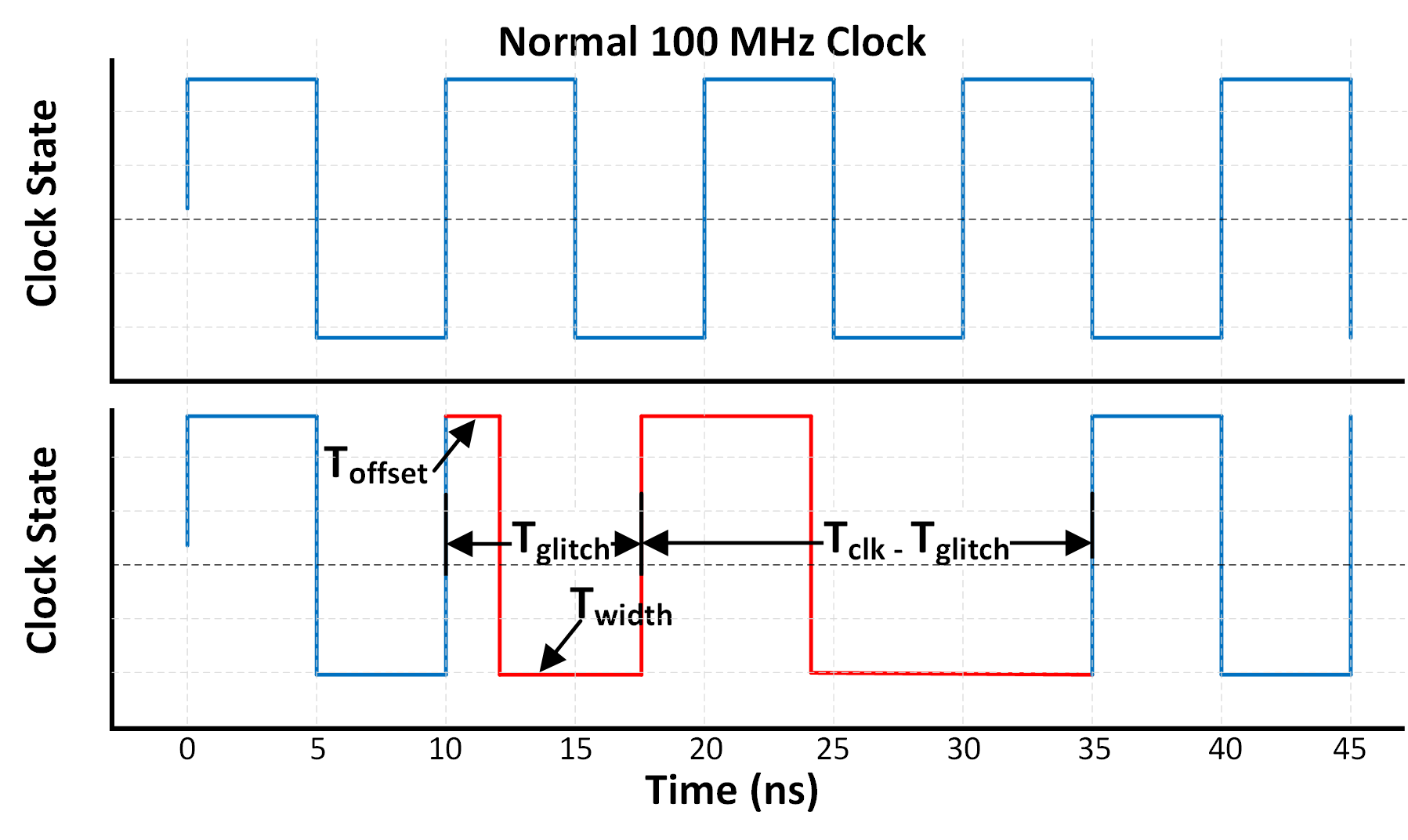}
	\vspace{-1.5em}
	\caption{Illustration of a $100$ MHz clock and a clock with a glitch. The top graph shows the regular clock signal with a stable period of $10$ ns, while the bottom graph illustrates a clock glitch, where the positive edge is advanced, resulting in a shortened clock period. This disruption can lead to timing violations and data corruption in digital circuits.}
    \Description{Illustration of a $100$ MHz clock and a clock with a glitch. The top graph shows the regular clock signal with a stable period of $10$ ns, while the bottom graph illustrates a clock glitch, where the positive edge is advanced, resulting in a shortened clock period. This disruption can lead to timing violations and data corruption in digital circuits.}
\vspace{-1.5em}
	\label{fig: Glitch}
\end{figure}

\vspace{-0.25em}
\subsection{Impact of Clock Glitch}
\vspace{-.15em}
A clock glitch attack deliberately and temporarily disrupts the clock signal, causing a misalignment between clock edges that corrupts the data being processed. In such an attack, an adversary introduces a brief pulse or delay into the clock signal, processing data earlier or later than intended. The resulting timing disturbance can lead to outcomes ranging from simple data misalignment to severe data corruption or even system control failures~\cite{yuce2015improving}.
Two parameters define the nature of a clock glitch: glitch offset and glitch width~\cite{yuce2015improving}. Collectively, these parameters determine how the glitch affects the system's timing and functionality~\cite{yuce2018fault}. 
\\\indent
\textbf{Glitch offset (T$_{offset}$)} defines the point in the clock cycle where the glitch is introduced, as shown in Figure~\ref{fig: Glitch}. A glitch occurring near the rising or falling edge of the clock can delay or advance transitions, misaligning data latching and clock edges. Depending on the timing, this disruption may affect different parts of the circuit, leading to data corruption in registers or incorrect state transitions.
\\\indent
\textbf{Glitch width (T$_{width}$)} describes the duration of the clock disruption (see Fig~\ref{fig: Glitch}). In a clock glitch attack, reducing the glitch width increases the intensity of the fault\footnote{``Fault intensity refers to the level of physical stress exerted on the microprocessor hardware, pushing it beyond its standard operating limits."~\cite{yuce2015improving}}.
The duration of the glitch influences how long the circuit remains unstable. This timing disturbance can lead to anything from minor data misalignment to severe data corruption or even a total failure of system controls. While longer glitches increase the likelihood of unintended behavior, not all glitch settings necessarily lead to data corruption.

\vspace{-1em}
\subsection{Threat Model}
\vspace{-0.25em}
We conduct FIA on an AI/ML system by inserting clock glitches during the inference process of an embedded neural network as our test case~\cite{eBNN}. While traditional glitching attacks have primarily targeted cryptographic systems and PIN verification, applying fault injection to AI/ML systems is a lucrative target because, unlike cryptographic implementations, AI/ML systems are particularly vulnerable due to their complex designs and noisy execution traces~\cite{gongye2024one, shuvo2023comprehensive}. Moreover, as the reliance on AI hardware accelerators grows, it introduces new security challenges and opportunities, not present in traditional cryptographic systems~\cite{yan2023defense}.

We follow the conventional assumptions in clock glitch attacks~\cite{chamelot2022sci, SYNFI, nasahl2023scramble, claudepierre2021traitor, yuce2018fault, yuce2015improving}. Specifically, we assume that the attacker has physical access to the circuit's clock signal. This access allows them to execute a \emph{single-glitch} attack with precise control over its width and offset. Using a phase-locked loop (PLL)~\cite{liu2020imperceptible}, the attacker can dynamically generate and modify glitch parameters, enabling precise manipulation of the clock signal during runtime. In the context of AI/ML applications, the attacker aims to inject glitches during critical operations to disrupt the program’s control flow, inducing data corruption or system misbehavior. This ultimately compromises model functionality and accuracy through carefully crafted perturbations. While we assume in-person manipulation, a sophisticated attacker may also tune these parameters remotely by updating FPGA firmware, further elevating the risk of such attacks.
\section{The Proposed Framework: CRAFT}\label{Sec: HW_design}
\vspace{-0.25em}
Prior works predominantly focused on post-silicon experiments, targeting applications, such as the final round of AES encryption, and used fault sensitivity analysis to evaluate cryptographic software vulnerabilities~\cite{yuce2015improving,yuce2018fault}. By
In contrast, we introduce CRAFT, a framework for systematically discovering and identifying the root causes of vulnerabilities in a RISC-V softcore processor. CRAFT achieves this by combining pre-silicon analysis with post-silicon validation, using targeted clock glitches. CRAFT assesses the effectiveness of these clock glitches on the RISC-V processor through a seven-step process, as shown in Figure~\ref{fig: Process}.
\vspace{-0.75em}
\subsection{Fault Injection Parameters} 

CRAFT begins by selecting the T$_{offset}$ and T$_{width}$ values informed by post-silicon PLL capabilities. This step aims to select coarse-grained parameters\footnote{Fine-tuning of these parameters occurs in the attack instrumentation step.} that induce timing violations, which are representative of potential fault scenarios, thereby reducing the search (post-silicon) space. The selection process is guided by post-silicon factors such as the onboard PLL capabilities, hardware constraints, operating conditions (temperature/voltage), clock skew, and clock jitter. The careful selection of these parameters helps achieve clock glitch effects such as pulse addition, cycle skipping, duty cycle change, and phase shift~\cite{marotta2024characterizing,velegalati2013glitch,he2024design}. By systematically exploring these parameters, CRAFT corrupts instruction(s) while avoiding system instability/crashes.
\vspace{-0.75em}\subsection{Attack Surface Exploration} CRAFT explores the attack surface by analyzing the target application to identify possible vulnerable instructions and execution points. To this end, CRAFT conducts static code analysis to locate sensitive instruction sequences, such as neural network inference layers, activation function computations, or matrix multiplication operations during model inference that translate to instructions such as jumps, branches, loads, and so on. Through this analysis, CRAFT generates a list of candidate instructions that represent the most promising targets for the fault injection. 
\vspace{-0.75em}\subsection{Fault Simulation} 
The pre-silicon simulation phase enables CRAFT to scrutinize parameters before reenacting them on hardware. Once CRAFT identifies instructions of interest, it begins fault injection to validate their susceptibility. CRAFT uses post-synthesis netlist simulations to test fault effects on the target instruction. CRAFT adjusts T$_{offset}$ and T$_{width}$ parameters within the simulation to identify all scenarios causing instruction corruption while avoiding system-wide failures (e.g., processor reset). This strategy helps CRAFT identify instructions exhibiting critical faults\footnote{During our experiments, we observed eight distinct categories of fault behavior. However, in this study, we focused only on critical faults.}.
\\\noindent\textbf{Critical fault.}\label{Critical_fault} CRAFT recognizes a fault as critical if it manifests in one of the following ways:
\begin{itemize}
\vspace{-0.25em}
\item \textbf{Instruction skip.} The instruction is skipped entirely or its result is corrupted (e.g., for a load instruction, its old value is retained, or a new corrupted value is stored in the destination register when the instruction retires).
\item \textbf{Program counter (PC) redirection.} The PC redirects to an incorrect address, disrupting program flow (e.g., for a jump instruction, a corruption in the next PC value computation leads to faults in subsequent PC calculations).
\end{itemize}
\vspace{-0.25em}
\textbf{Non-critical fault.}  Clock glitches can freeze the CPU, suspending it in an undefined state~\cite{Freeze}. Recovery from such states often requires a CPU reset. While harmful (denial of service), such crashes are less useful for stealthy exploitation, as they are immediately noticeable and do not directly provide the attacker with stealthy control or knowledge. CRAFT classifies these faults as non-critical due to limited application and does not explore them further.

\vspace{-.75em}\subsection{Risk Assessment Table} After the simulation phase, CRAFT compiles a risk assessment table (RAT) to prioritize the most critical vulnerabilities. This table helps guide efficient and targeted attacks by capturing the following attributes(in percentage):

\begin{itemize}
\vspace{-0.25em}
    \item \textbf{Instruction Type.} The category of the instruction (e.g., arithmetic, logic, control flow).
    \item \textbf{Priority Score.} A percentage score that represents the number of instructions that exhibited critical faults among the total number of instructions evaluated.
\end{itemize}
\vspace{-0.25em}
The RAT helps CRAFT prioritize instructions for targeted post-silicon fault injection. By quantifying fault reproducibility, RAT enables CRAFT to focus on instructions with high impact and reproducibility. This makes the RAT a valuable utility for two key groups: \textit{attackers}, who can pinpoint vulnerable instructions at specific pipeline stages, and \textit{system designers}, who can proactively address these vulnerabilities to enhance the processor defenses.


\vspace{-.75em}\subsection{Attack Instrumentation}\vspace{-0.3em} Insights from pre-silicon simulations (captured in the RAT) guide the selection of glitch parameters for targeted and efficient post-silicon experiments. CRAFT replicates these parameters to execute fault injections and monitors the system behavior using an on-chip debugger. An attack is considered successful when it produces the expected fault outcomes, such as instruction skips, corrupted memory values (e.g., failed loads), or unauthorized PC redirection.
\vspace{-.75em}\subsection{Root-Cause Analysis}\vspace{-0.3em}
CRAFT is used to conduct a root-cause analysis to understand why specific instructions fail in the presence of a fault. This analysis helps trace the fault's origin to underlying microarchitectural events, such as shared pipeline registers or instruction decoding errors. Identifying patterns in the failing instructions then guides the refinement of glitch parameters. When recurring failures are observed, the specific glitch width or offset are flagged as the fault's trigger. Ultimately, this phase also provides insights into potential countermeasures for mitigating the identified vulnerabilities.


\vspace{-.75em}\subsection{Attack Validation} \vspace{-0.3em}
Finally, CRAFT validates its pre-silicon attack results by repeating the experiments on actual hardware and comparing the outcomes. This process confirms that the vulnerabilities identified in the simulated environment (pre-silicon) also exist in real-world conditions (post-silicon), providing empirical evidence that the target is vulnerable in both pre- and post-silicon stages.

\section{Case Study: Targeting AI/ML Applications}\label{FIA_Method}
We now demonstrate CRAFT's effectiveness with a case study on AI/ML applications.
\vspace{-0.75em}
\subsection{Fault Injection Parameters}\label{FIA_Params}\vspace{-0.25em}
We used Vivado $2020.2$ for pre-silicon instrumentation, and deployed the (same) cv$32$e$40$x soft-core processor on the Xilinx Kintex-$7$ XC$7$K$160$T FPGA (SAKURA-X) for post-silicon verification.  To demonstrate CRAFT's effectiveness, we targeted an embedded neural network inference application~\cite{eBNN}. CRAFT tested $17\times17$  distinct clock glitch configurations, varying  glitch offset (T$_{offset}$) and  glitch width (T$_{width}$) between $0.278$ns and $8.89$ns, with a step size of $0.5$ns\footnote{We set T$_{offset}$ and T$_{width}$ to align with the $20$ns clock period for our $50$MHz design. This precise timing enables us to detect violations in fault scenarios. We chose $0.5$ns for these parameters, considering post-silicon PLL capabilities, signal integrity, hardware limitations, operating conditions (temperature and voltage), clock skew, and jitter. This careful selection helps us observe effects such as pulse addition, cycle skipping, duty cycle changes, and phase shifts.~\cite{marotta2024characterizing,velegalati2013glitch,he2024design}.}.
This configuration was broad enough to induce various timing violations---covering a number of fault scenarios---while maintaining efficiency during multiple fault injection experiments. 
\vspace{-.75em}\subsection{Attack Surface Exploration}\label{Exploration}\vspace{-0.25em}
We selected eight instructions crucial for neural network inference, following well-established paradigms~\cite{kazemi2021depth,faultdetective}. We compiled the code and analyzed eight carefully chosen instructions critical to the inference process, such as the loop-trip count check (a branch instruction in RISC-V) and memory load/store operations for the weights or biases. We focused on this subset for workload relevance, to ensure coverage of different instruction formats and pipeline behaviors. \textbf{However, CRAFT is \emph{not} limited to these eight instructions}; it is scalable and capable of evaluating all instructions. We focused on a subset of instructions to manage experimental complexity; however, no fundamental limitation prevents analyzing more RISC-V ISA instructions except time and resources.
\begin{table}[t]
\setlength{\tabcolsep}{16pt}
\centering
\small 
\caption{Risk assessment table (RAT) displaying the total percentage of faults observed across the four pipeline stages of the RISC-V processor, with the sum of all cells equaling 100\%.}
\vspace{-0.75em}
\label{tab:RAT}
    \begin{tabular}{|c|c|c|c|} \hline
    \multirow{2}{*}{\textbf{Instruction}} & \multicolumn{3}{c|}{\textbf{Total Faults Observed (\%)}}                                                      \\ \cline{2-4} 
                                      & \multicolumn{1}{l|}{\textbf{IF/ID}} & \multicolumn{1}{l|}{\textbf{ID/EX}} & \multicolumn{1}{l|}{\textbf{EX/WB}} \\ \hline
    \textbf{\texttt{c.addi}} & 0.40  & 5.24 & 6.45 \\ \hline
    \textbf{\texttt{auipc}}  & 0.0   & 0.0  & 2.03 \\ \hline
    \textbf{\texttt{JAL}}    & \textbf{14.92} & \textbf{13.31}& \textbf{6.85} \\ \hline
    \textbf{\texttt{bne}}    & 2.42  & 1.61 & 3.23  \\ \hline
    \textbf{\texttt{bge}}    & 1.61  & 6.05 & 1.21  \\ \hline
    \textbf{\texttt{c.lwsp}} & 1.61  & 0.81 & 2.82  \\ \hline
    \textbf{\texttt{c.mv}}   & 0.0   & 0.81 & 0.0   \\ \hline
    \textbf{\texttt{lw}}     & \textbf{11.29} & \textbf{9.27} & \textbf{8.06}  \\ \hline
\end{tabular}
\vspace{-1.5em}
\end{table}

\vspace{-.75em}\subsection{Fault Simulation}\label{Fault__Simulation}\vspace{-0.25em}
For the selected eight instructions, CRAFT tested $9248$ glitch configurations ($17$ glitch offsets \(\times\) $17$ glitch widths \(\times\) $4$ pipeline stages \(\times\) $8$ instructions) to exhaustively identify timing paths that trigger critical faults. CRAFT inserts a single-clock glitch at the precise moment to disrupt the critical timing path of a targeted instruction, thereby inducing execution errors. CRAFT targets each instruction as it moves through individual pipeline stages to achieve three goals: (1) mapping the timing path associated with each instruction at each stage, (2) assessing the instruction’s vulnerability within each stage, and (3) narrowing the post-silicon validation space by shortlisting \emph{only} those glitch parameters that produce critical faults.

\vspace{-.75em}\subsection{Risk Assessment Table}\label{R_A_T}\vspace{-0.25em}
Table~\ref{tab:RAT} lists the eight selected instructions and categorizes the critical faults observed in each pipeline stage. The table displays the total percentage of critical faults observed, with a higher percentage indicating greater vulnerability to clock glitches\footnote{Since in RISC-V processor adjacent pipeline stages share the pipeline registers, we show faults as a cumulative sum for each shared stage: IF/ID, ID/EX, and EX/WB.}. We evaluated only these eight instructions, following the well-established paradigm in academic literature~\cite{kazemi2021depth,faultdetective}. However, this is \emph{not} a limitation because CRAFT is designed to be scalable, enabling it to evaluate any instruction. Using the Table~\ref{tab:RAT} RAT ranking, we hypothesized the causes of faults and validated these hypotheses through our systematic approach, which is detailed in the following subsections.

\vspace{-.75em}\subsection{Attack Instrumentaion}\label{Attack__Instrumentaion}\vspace{-0.25em}
In the pre-silicon phases (Sections~\ref{Fault__Simulation} and~\ref{R_A_T}), CRAFT began with $9248$ glitch configurations. These phases guide CRAFT to determine (a) the exact timing path of each instruction to induce a timing violation, (b) which instruction(s) is vulnerable and at which pipeline stage, and (c) the impact of glitches on these instructions. Using the insights from these pre-silicon phases, CRAFT identified and narrowed down $248$ cases that resulted in critical faults, reducing our focus (for the post-silicon) from $9248$ to $248$ configurations—achieving a $97.31$\% reduction.
\begin{figure}[t]
\centering
\vspace{-1.0em}
         \includegraphics[scale =0.6]{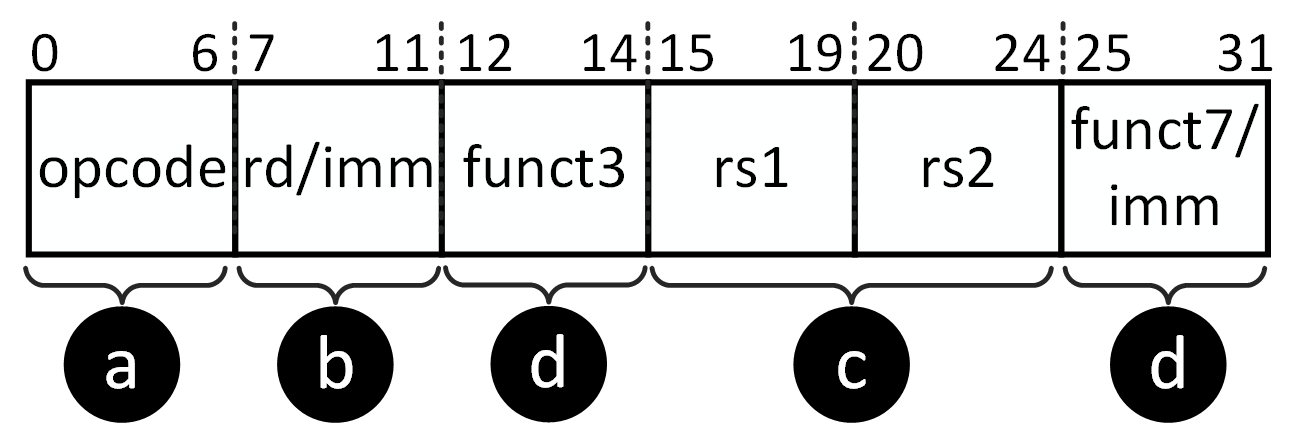}
\vspace{-1.0em}
\caption{RISC-V instruction format depicting the opcode,  register fields, function codes, and immediate values, highlighting the utilization of various fields depending on the instruction type, \textit{e.g.,} R-type, I-type. Corruption in different fields of instruction can lead to distinct behaviors.}
\Description{RISC-V instruction format depicting the opcode,  register fields, function codes, and immediate values, highlighting the utilization of various fields depending on the instruction type, \textit{e.g.,} R-type, I-type. Corruption in different fields of instruction can lead to distinct behaviors.}
\vspace{-1.5em}
	\label{fig: Instr_format}
\end{figure}

\begin{figure}[b]
\vspace{-1.9em}
	\includegraphics[width=0.49\textwidth]{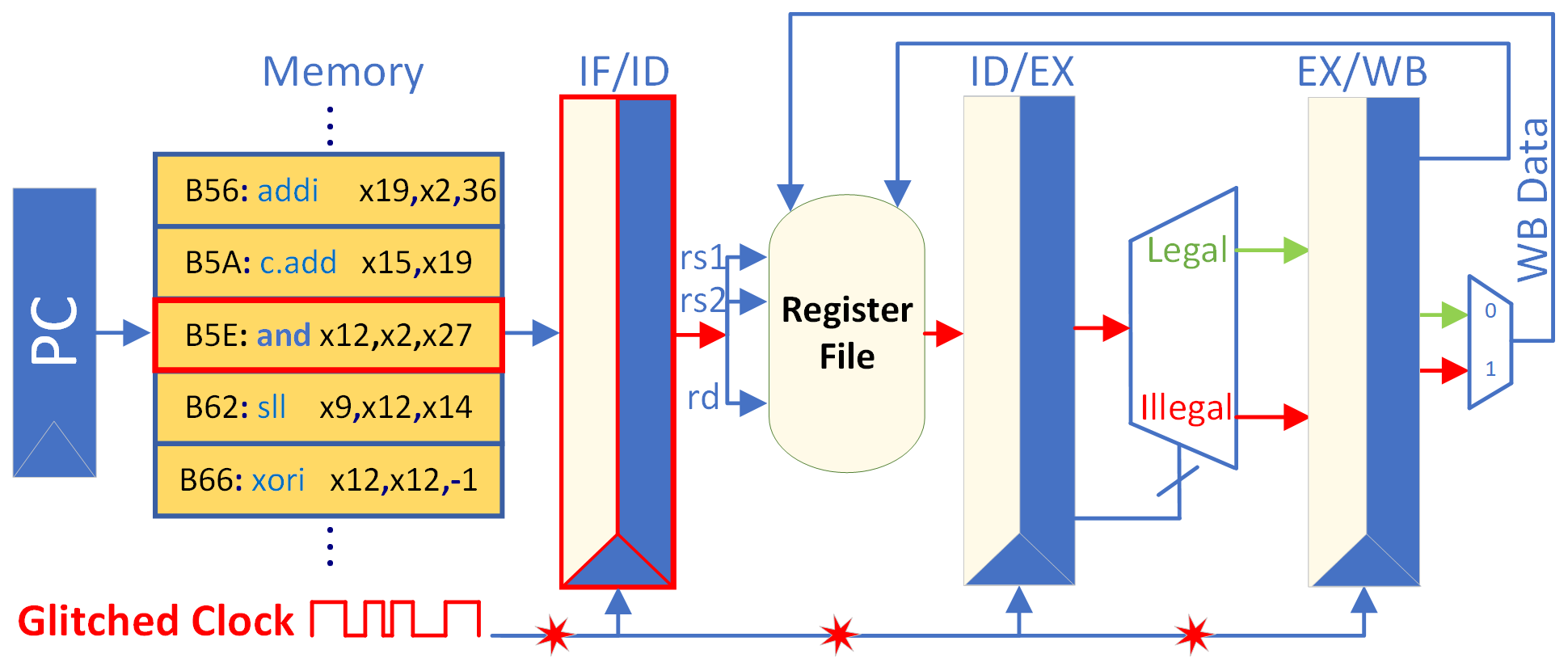}
	\vspace{-2em}
	\caption{Visualization of the cv$32$e$40$x $4$-stage pipeline processor. In the Fetch stage, when the program counter loads the `and' instruction from memory, a clock glitch disrupts the operation, corrupting the pipeline registers shared between the Fetch and Decode stages, resulting in the misclassification of the legal `and' instruction as an illegal instruction.}
    \Description{Visualization of the cv$32$e$40$x $4$-stage pipeline processor. In the Fetch stage, when the program counter loads the `and' instruction from memory, a clock glitch disrupts the operation, corrupting the pipeline registers shared between the Fetch and Decode stages, resulting in the misclassification of the legal `and' instruction as an illegal instruction.}
	\vspace{-2em}
	\label{fig: Pipeline}
\end{figure}



\vspace{-.75em}
\subsection{Root-Cause Analysis}\label{RootCause}
To effectively defend against identified faults, it is essential to understand their root cause(s). Therefore, we conducted an in-depth analysis of these faults, starting by examining them with the assistance of the RISC-V instruction set manual~\cite{waterman2014risc}. Figure~\ref{fig: Instr_format} shows the instruction format of the RISC-V ISA. The format consists of an opcode, a destination register, source registers, and, depending on the instruction type, either a sub-function (for R-type instructions) or an immediate (imm) field (for I-type instructions). A clock glitch that corrupts this format can have four possible outcomes: \encircle{a} altering the opcode to execute an unintended instruction, causing incorrect system behavior; \encircle{b} corrupting the destination register, leading to incorrect data storage; \encircle{c} corrupting source registers, causing data to be fetched from incorrect locations; or \encircle{d} affecting the sub-function or immediate fields, resulting in incorrect operation execution or erroneous PC calculations. 

To better understand this scenario, consider an example involving the RISC-V processor. Figure~\ref{fig: Pipeline} illustrates the RISC-V processor pipeline suffering from a clock glitch attack.
\begin{figure}[t]
\vspace*{-1.5em}
\centering
         \includegraphics[width=0.48\textwidth]{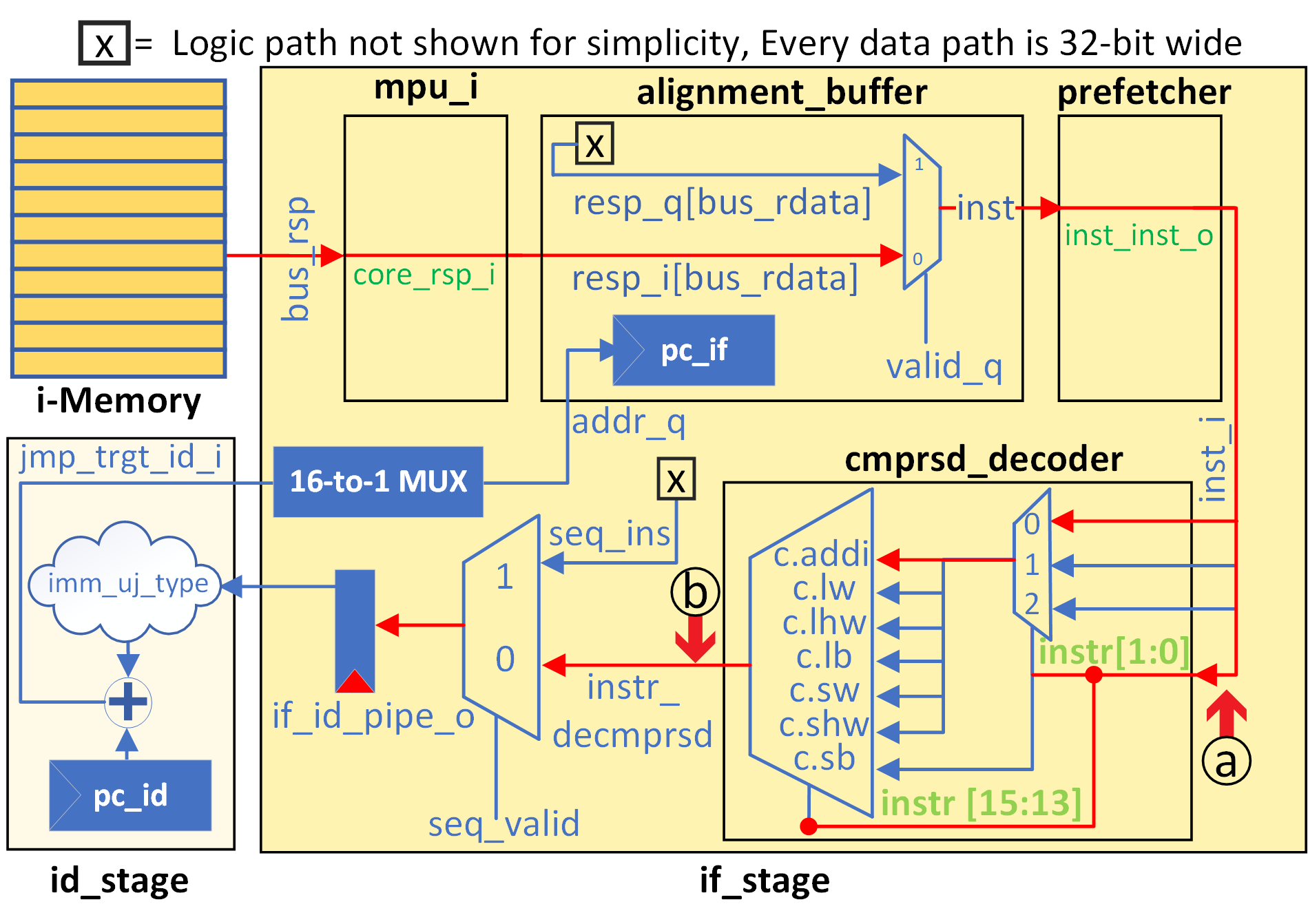}
\vspace{-2.25em}
	\caption{The root cause of failure in the cv$32$e$40$x processor. A clock glitch disrupts the long combinational logic path, causing the input to the compressed decoder module to be latched prematurely (see \textcircled{a}). This leads to misclassifying legal instructions as illegal in subsequent clock cycles.}
    	\Description{The root cause of failure in the cv$32$e$40$x processor. A clock glitch disrupts the long combinational logic path, causing the input to the compressed decoder module to be latched prematurely (see \textcircled{a}). This leads to misclassifying legal instructions as illegal in subsequent clock cycles.}
 
\vspace{-1.5em}
	\label{fig: Critical_Path}
\end{figure}
At every clock cycle, the PC points to the memory address containing the next instruction to fetch. The fetched instruction is stored in the IF/ID pipeline register\footnote{In the RISC-V source code, the IF/ID pipeline register is named \texttt{$if\_id\_pipe\_o$}. In this work, we alternatively refer to it by the same name to maintain consistency.\label{regname}} which is shared between the fetch (IF) and the decode (ID) stage (shown as IF/ID in Figure~\ref{fig: Pipeline}). Consider the scenario where the PC reaches the value of 0xB5E, and an `\texttt{AND}' instruction is fetched. A precise clock glitch corrupts the instruction stored in the IF/ID pipeline register---our technique in Section~\ref{FIA_Method} has identified this as an instruction that can be corrupted between the IF and ID stage without corrupting other pipeline stages. 
\\\indent
Once an instruction reaches the decode-execute (ID/EX) junction, if the instruction conforms to the RISC-V instruction format rules, it proceeds to further execution; otherwise, it is flushed from the pipeline. This flush skips the `illegal' instruction (newly corrupted), and the PC advances to the next valid instruction. The RISC-V ISA classifies such invalid instructions as `illegal.' In this example, a clock glitch corrupted the fetched `\texttt{AND}'  instruction, causing its misclassification as `illegal' and subsequent skip, thus preventing any value from being written to its destination register.
\\\indent
The RISC-V ISA manual specifies conditions for flagging illegal instructions but offers no explicit corrective actions, placing the burden of exception handling on the user. Our analysis of the RISC-V source code showed that the inserted clock glitch corrupted the IF/ID pipeline register\textsuperscript{~\ref{regname}}. This register, named \texttt{$if\_id\_pipe\_o$} in the source code, is driven by a long combinational path. The glitch caused a timing violation, corrupting its contents. The \texttt{$if\_id\_pipe\_o$} register holds the fetched instruction and the control signals required for decoding the instruction in the subsequent ID stage. Therefore, its contents are vital to ensuring proper instruction flow and maintaining synchronization within the pipeline.
\\\indent
Recognizing this register's critical role, let us examine how a fault affects the most vulnerable instruction identified by the RAT: jump (`\texttt{JAL}'), with the aid of Figure~\ref{fig: Critical_Path}, which illustrates the critical path identified through pre-silicon testing and source code analysis. 
`\texttt{JAL}' computes the new target PC by summing the current PC value of the decode stage (`pc\_id') with an immediate value (`imm\_uj\_type') provided by the \texttt{$if\_id\_pipe\_o$} register. 
The \texttt{$if\_id\_pipe\_o$} register derives its input from the `compressed decoder' module, which 
handles RISC-V's compressed instruction set extension, converting $16$-bit compressed instructions into their $32$-bit equivalents.\\\indent A clock glitch corrupts the input to the compressed decoder, which stores an incorrectly formatted value in \texttt{$if\_id\_pipe\_o$}. This corruption alters the `imm\_u\_type' field and causes the processor to compute the PC incorrectly, jumping to an invalid address. The processor then classifies the resulting instruction as illegal. Without an exception handler, it skips the instruction to continue execution. Our root-cause analysis identifies the IF/ID pipeline register and its preceding combinational logic as the sources of fault injection vulnerabilities. Glitches cause the register to latch incorrect instruction data, leading to control-flow errors (e.g., wrong PC) or data corruption (e.g., incorrect register writes) without triggering system-level alerts. 


\begin{table}[t]
\centering
\vspace{-1.0em}
\large
\caption{{Experimental results showing various glitch offset and width configurations that lead to output corruption. Adjusting the glitch width produces effects such as instruction skipping, loading all zeroes, or selective corruption of the MSBs.}}
\label{tab:Parameters}
\vspace{-0.75em}
\resizebox{\columnwidth}{!}{%
\begin{tabular}{|c|c|c|c|c|c|}
\hline
\textbf{\begin{tabular}[c]{@{}c@{}}Case\\ \#\end{tabular}} &
  \begin{tabular}[c]{@{}c@{}}$\textbf{T}_{offset}$\\  $(ns)$\end{tabular} &
  \begin{tabular}[c]{@{}c@{}}$\textbf{T}_{width}$ \\  $(ns)$\end{tabular} &
  \begin{tabular}[c]{@{}c@{}}$\textbf{T}_{glitch}$\\  $(ns)$\end{tabular} &
  \textbf{\begin{tabular}[c]{@{}c@{}}Illegal Flag \\ Raised\end{tabular}} &
  \textbf{\begin{tabular}[c]{@{}c@{}}Effect \\ Observed\end{tabular}} \\ \hline
\textbf{1} & \multirow{4}{*}{0.833} & $\leq$2.967    & $\leq$3.8      & \cmark & Instruction Skip        \\ \cline{1-1} \cline{3-6} 
\textbf{2} &                        & 3.067 -- 3.567 & 3.901 -- 4.400 & \cmark & Data Zeroization        \\ \cline{1-1} \cline{3-6} 
\textbf{3} &                        & 3.667 -- 4.289 & 4.504 -- 5.121 & \xmark & Data Zeroization        \\ \cline{1-1} \cline{3-6} 
\textbf{4} &                        & 4.289 -- 4.339 & 5.112 -- 5.172 & \xmark & Partial Data Corruption \\ \hline
\end{tabular}
}
\vspace{-1.25em}

\end{table}
\vspace{-.5em}
\subsection{Update After Attack Instrumentation}\label{Crafting}
As shown in Figure~\ref{fig: Process}, after Step 5, the `Update' step can be utilized to fine-tune fault parameters. Building on insights from Sections~\ref{Sec: HW_design} and~\ref{FIA_Method}, CRAFT first identified the timing parameters required to induce timing violations and the critical paths vulnerable to clock glitches through root-cause analysis. These insights were then applied using the `Update' step to reduce the (post-silicon) search space and focus on high-impact fault scenarios. Furthermore, RAT helped reveal that the load (`\texttt{lw}') and jump (`\texttt{JAL}') instructions were the most susceptible, each producing multiple critical faults. This also correlates with their complexity—`\texttt{lw}' performs memory access and write-back while `\texttt{JAL}' updates the PC. Both require multi-step operations across the pipeline and rely on timing-critical computations \textit{e.g.,} memory fetch, adder for PC. By contrast, simpler instructions like adding upper immediate to PC (\texttt{auipc}) and move (\texttt{c.mv}) produced few or no faults in most stages, reflecting that some operations have more timing slack or simpler logic.

Using RAT as an initial guide, we selected the two most vulnerable instructions, `\texttt{\texttt{lw}}' and `\texttt{JAL}', for fine-grained tuning. We (correctly) hypothesized that focusing on these would reveal precise glitch settings, producing new fault behaviors that were missed in our initial sweep. We concentrated on their most vulnerable point as identified in the root-cause analysis—the IF/ID pipeline stage transition when instructions latch into \texttt{$if\_id\_pipe\_o$} register. We fine-tuned glitch settings around known fault-inducing parameters \textit{e.g.,} when a $3.0$ns T$_{offset}$ with $4.0$ns T$_{width}$ caused a fault, we tested nearby values ($2.9$ns, $3.1$ns) to uncover subtle behaviors. This targeted approach, though more manual than our broad sweep, was guided by known weak spots, making it manageable and effective.

By tuning these parameters, we discovered two additional ranges where an attacker could utilize CRAFT's insights to either skip specific instructions or redirect the program flow \textbf{without raising an `illegal' instruction flag} (refer Section~\ref{Sec:Results} for details on the parameters). This can allow sophisticated attackers to cause significant disruptions while leaving the processor unaware of the corruption.
The expected outcomes of such an attack include:
\begin{itemize}
    \item \textbf{Case \#1:} The instruction is skipped, triggering the `illegal' flag and the exception handler.
    \item \textbf{Case \#2:} The destination register (`rd') is zeroed while triggering the `illegal' flag and the exception handler.
    \item \textbf{Case \#3:} The destination reg (`rd') is zeroed \textbf{without} triggering the `illegal' flag or the exception handler.
    \item \textbf{Case \#4:} The destination reg (`rd') is partially corrupted \textbf{without} triggering the `illegal' flag or exception handler.
\end{itemize}
In \textbf{Case \#1} and \textbf{Case \#2}, the `illegal' flag triggers an exception, redirecting the program to the exception handler address. Without a defined handler, the program skips the instruction and continues execution. The `illegal' flag indicates the glitch caused the core to latch an invalid instruction format when it captured the decoder output too early. These cases differ in register write-back timing: \textbf{Case \#2} writes a zero value before the exception takes effect, while \textbf{Case \#1} skips the instruction completely without writing. The core detects both problems by raising exceptions.
\\\indent
By contrast, the `illegal' instruction flag is not even triggered in \textbf{Case \#3} and \textbf{Case \#4}. This is because in these cases, the \textit{input} to the `compressed decoder' is latched just in time, avoiding detection as an `illegal' instruction (refer \textcircled{a} in Figure~\ref{fig: Critical_Path}). However, the \textit{output} of the `compressed decoder' still fails to meet timing constraints and does not update in time (refer \textcircled{b} in Figure~\ref{fig: Critical_Path}). As a result, the `compressed decoder' processes what seems to be valid data but actually executes corrupted information. Table~\ref{tab:Parameters} shows glitch parameter ranges that lead to each of these cases\footnote{Migrating to a different device may require a one-time re-run of the fault simulation (post-synthesis) step to account for variations, such as changes in netlist and logic delay.}. 
\vspace{-0.15em}
\begin{figure}[t]
\centering
\vspace{-0.95em}
 \includegraphics[width =0.48\textwidth]{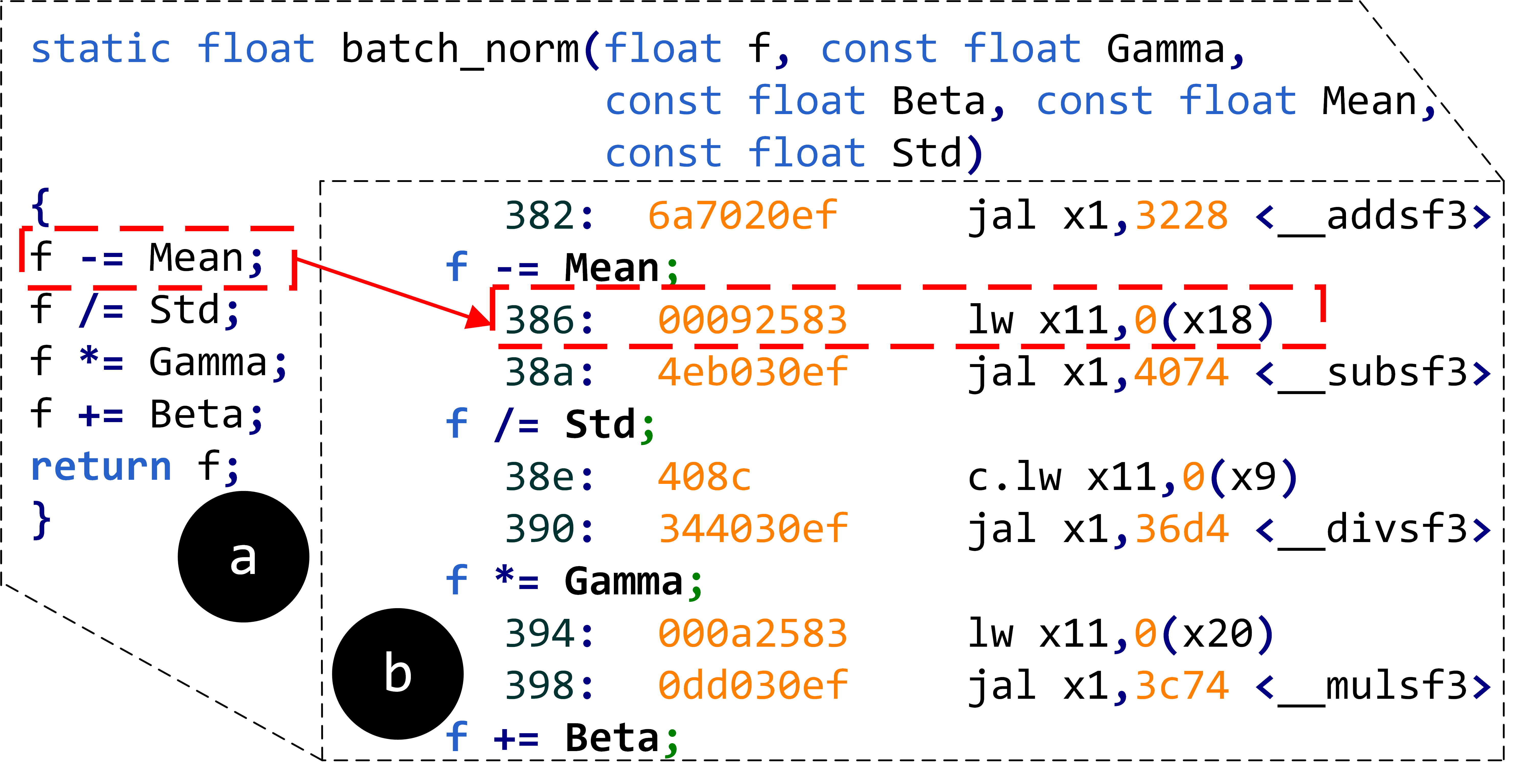}
\vspace{-2.0em}
	\caption{(a) C-code for MNIST inference and (b) the targeted `load' instruction in the assembly by CRAFT }
    \Description{(a) C-code for MNIST inference and (b) the targeted `load' instruction in the assembly by CRAFT. }
 
\vspace{-1.75em}
	\label{fig: code}
\end{figure}
\begin{figure*}[t]
\centering
         \includegraphics[width =\textwidth]{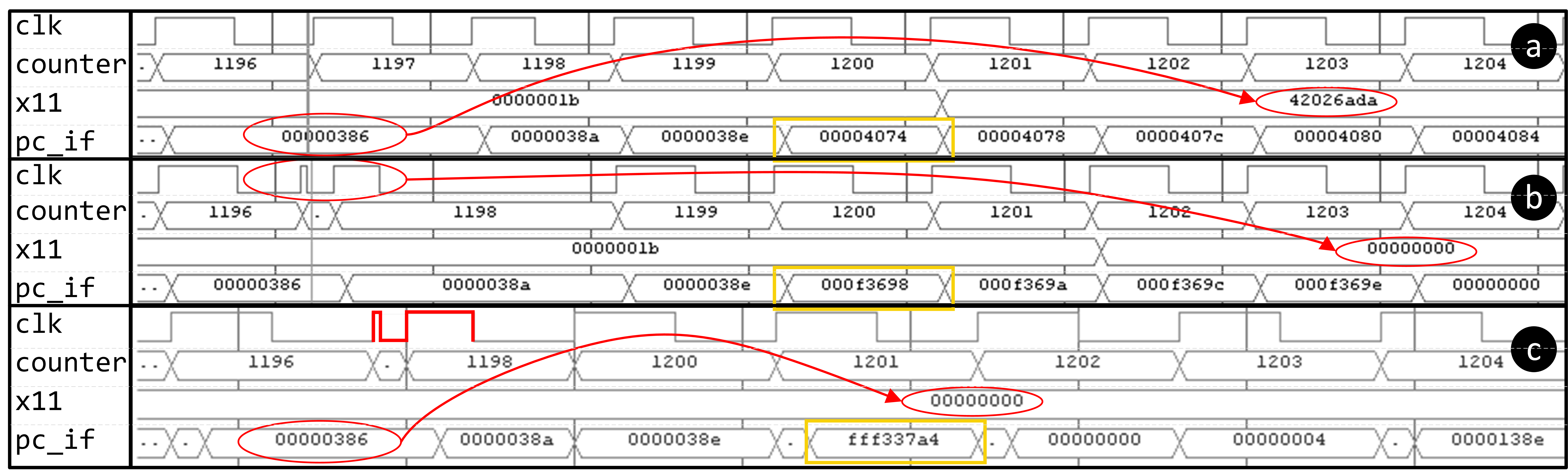}
\vspace{-2.0em}
	\caption{CRAFT targetting the RISC-V ISA in a neural network inference algorithm~\cite{eBNN} is depicted in three scenarios: (a) \emph{without} and \emph{with} a clock glitch in (b) pre- and (c) post-silicon. The first-order effect of the clock glitch attack corrupts the destination register `x11' associated with the `\texttt{lw}' instruction, loading all zeroes. The second-order effect corrupts the immediate field of the `\texttt{JAL}' instruction, disrupting the normal program flow and causing incorrect control flow redirection. In (c), the glitched clock signal is drawn by hand due to the limited sampling capacity of the integrated logic analyzer (ILA).}
    	\Description{CRAFT targetting the RISC-V ISA in a neural network inference algorithm~\cite{eBNN} is depicted in three scenarios: (a) \emph{without} and \emph{with} a clock glitch in (b) pre- and (c) post-silicon. The first-order effect of the clock glitch attack corrupts the destination register `x11' associated with the `\texttt{lw}' instruction, loading all zeroes. The second-order effect corrupts the immediate field of the `\texttt{JAL}' instruction, disrupting the normal program flow and causing incorrect control flow redirection. In (c), the glitched clock signal is drawn by hand due to the limited sampling capacity of the integrated logic analyzer (ILA).} 
\vspace{-1.3em}	
\label{fig: load}
\end{figure*}

CRAFT illustrates how adjusting the glitch width can toggle between raising an illegal exception or bypassing it entirely. The implications of these fault cases are serious. A silent skip (\textbf{Case \#3} or \textbf{Case \#4} affecting a branch or jump) could let an attacker bypass a security check without any log or trace—\textbf{leaving the processor completely unaware of the fault}. Silent data corruption in a load (\texttt{lw}) instruction could alter a critical parameter in an algorithm, such as modifying the weights in a neural network or corrupting the input data used for training. Such vulnerabilities pose significant risks to critical applications, including neural networks. A well-timed attack could silently corrupt crucial operations, such as loading weights or biases from memory, potentially leading to undetectable critical misclassifications.

\section{Attack Validation: Pre- and Post-Silicon} \label{Sec:Results}
\vspace{-0.25em}

\noindent\textbf{Test-Platform.} We used Vivado $2020.2$ and the cv$32$e$40$x soft-core processor for the pre-silicon instrumentation. As part of our case study, we explored the impact of clock glitch-based fault injections on a neural network algorithm running inferences for the MNIST dataset~\cite{eBNN}.
We ran the RISC-V processor at $50$MHz with a timing constraint of $20ns$ specified in the constraint (*.xdc) file. We deployed the cv$32$e$40$x soft-core processor on the Xilinx Kintex-$7$ XC$7$K$160$T FPGA (SAKURA-X) for post-silicon verification. Using the RISC-V compiler toolchain we generated the binary file of the inference code, comprising the necessary machine instructions and associated data. A custom controller module was developed to enable data transmission from the host PC to the FPGA. 


We used the FPGA's internal PLL to introduce glitches to the clock signal. When the processor core detects the target instruction, it relays a trigger signal to the glitch circuitry. The PLL generates three distinct clock signals: `CLK0', which serves as the reference base clock at $50$ MHz, `CLK1' and `CLK2', which are phase-shifted versions of `CLK0'. We can modulate the glitch parameters by dynamically reconfiguring the phase difference among these clocks. The phase difference of `CLK1' relative to `CLK0' and `CLK2' defines T$_{offset}$ and T$_{width}$, respectively. To verify the effects of clock glitches, we integrated the internal logic analyzer (ILA) into our design to capture, observe, and analyze core signals, such as, the PC value, register file contents, and critical control signals. The ILA monitored and sampled these signals at $300$ MHz frequency~\footnote{Beyond this frequency, the signals and values start to exhibit unreliability, compromising accurate observations.}.
\vspace{-0.75em}
\subsection{Pre-Silicon Verification}
\vspace{-0.25em}
We used gate-level post-synthesis netlist to orchestrate CRAFT and inject faults for two reasons: (i) it is the earliest design stage that includes the circuit's timing information. Analysis at this stage allows for identifying potential vulnerabilities and timing-related issues that could be challenging and costly to address later in the design process, and (ii) it enables rapid and detailed analysis with minimal overhead compared to post-silicon evaluations. 

Figure~\ref{fig: code}~(a) and (b) present a snippet of the C code and the corresponding assembly code of the inference algorithm, respectively.
The attack target is the `\texttt{lw}' instruction with a PC value of $0$x$386$.
CRAFT induced timing violations by injecting glitches into the nominal clock, adjusting T$_{offset}$ and T$_{width}$ to ensure that the data does not reach the \texttt{$if\_id\_pipe\_o$} register in time for the next clock edge.
Figure~\ref{fig: load} (a) shows fault-free execution of the neural network inference model (without clock glitch).
A $64$-bit counter increments with each rising clock edge to mark time progression. At clock cycle $1197$, the `\texttt{lw}' instruction is fetched with a PC value of $0$x$386$. In a fault-free  execution, 
the instruction completes when the counter reaches $1201$, loading the value $0$x$42026ada$ into the register `x11', as shown in red.



Figure~\ref{fig: load} (b) shows the first-order effect of the clock glitch (marked in red ellipses), showing zeroization of register `x11' associated with the `\texttt{lw}' instruction as expected in \textbf{Case \#2} and \textbf{Case \#3}. It also depicts the second-order effect (marked in yellow rectangles), where the immediate value in the `\texttt{JAL}' instruction is corrupted, resulting in incorrect `pc\_if' redirection. Normally, `\texttt{JAL}' correctly redirects `pc\_if' to offset $0$x$4074$, as shown in Figure~\ref{fig: load} (a). Through CRAFT, however, `pc\_if' is incorrectly redirected to $0$xF$3698$ due to premature latching in the `compressed decoder' module, as explained in Section~\ref{Crafting}, causing cascading faults in the program flow.

\vspace{-0.75em}
\subsection{Post-Silicon Verification}
\vspace{-0.25em}

We were able to successfully reproduce similar effects at the post-silicon level for \textbf{Case \#1}, \textbf{Case \#3}, and \textbf{Case \#4}. Figure~\ref{fig: load} (c) illustrates the verification results for \textbf{Case \#1}, where the attacked load instruction at the PC value $0$x$386$ is skipped (highlighted in red ellipse), raising the illegal instruction flag. The glitched clock signal is drawn by hand due to the limited sampling capacity of the ILA. The glitched clock experiences rapid transitions from $low \leftrightarrow high$ within $0.833$ns, significantly shorter than the ILA’s sampling period of $3.33$ns, resulting in undersampling of the glitch transitions. 
\\\indent
For \textbf{Case \#2}, the fault effects differ between pre-silicon and post-silicon experiments. The `x11' register loaded non-zero values during post-silicon verification versus all zeroes in pre-silicon. 
We speculate that the clock glitch that the prematurely latched input to the `compressed decoder' differs between pre-silicon and post-silicon results. This input consisting of the fetched instruction dictates the final value written to the `x11' register as explained in Section~\ref{RootCause}. Also, the fetched PC gets redirected to $0$xFFF$337$A$4$ instead of $0$xF$3698$, as shown in Figure \ref{fig: load} (b). 
Although the glitch effects are similar under the same settings, we attribute the differences in corrupted values to two factors: (1) critical path delay variations between post-synthesis and place-and-route netlists, and (2) the absence of the fine-grained precision for glitch generation—available in pre-silicon—due to the FPGA's clocking constraints in post-silicon, which made it difficult to replicate the exact glitch settings.
We confirmed the existence of silent faults for \textbf{Case \#3} and \textbf{Case \#4} in post-silicon.

\vspace{-0.5em}

\section{Discussions} \label{Sec:Discussions}
\vspace{-0.25em}
This section examines CRAFT's implications for RISC-V systems and evaluates potential security countermeasures.
\vspace{-0.35em}
\subsection{Further Implications of CRAFT}
\vspace{-0.25em}
While this study discloses a specific set of vulnerabilities, it is not confined to a single instruction or ISA. In the RISC-V architecture, different immediate fields play crucial roles, \textit{e.g.,} `imm\_i\_type' is used for load and arithmetic instructions, `imm\_s\_type' encodes store offsets, `imm\_sb\_type' is used for branch jump targets, and `imm\_u\_type' represents larger constants or address segments. All of these immediate types depend on the \texttt{$if\_id\_pipe\_o$}pipeline register, making them \textit{potentially} susceptible to CRAFT.


\vspace{-0.35em}
\subsection{Scalability}\label{Scalability}
\vspace{-0.25em}
This study evaluates eight instructions following the well-established paradigm in the academic literature~\cite{kazemi2021depth,faultdetective}; however, the proposed methodology can be readily scaled to cover the entire RISC-V instruction, providing broader applicability and more comprehensive insights into mitigating vulnerabilities. No fundamental limitation prevents the analysis from being carried out on more  RISC-V instructions beyond the investment of time and resources.

\vspace{-0.35em}
\subsection{Genericness}
\vspace{-0.25em}
Our results highlight the effectiveness of CRAFT's as a versatile fault analysis framework. We chose neural network inference as a test case because it represents a real-world, security-critical application where both timing accuracy and data integrity are vital. Analyzing fault effects in this context revealed complex vulnerabilities that could be exploited in practical scenarios. However, the methodology underlying CRAFT is \emph{not}  restricted to this specific workload. While CRAFT focuses on hardware-level vulnerabilities, it is adaptable to a wide range of workloads and instruction sets.


\vspace{-0.35em}
\subsection{Potential Countermeasures}
\vspace{-0.25em}
Although generic countermeasures, such as spatiotemporal redundancy, can help address the identified vulnerabilities, their broad implementation can result in substantial overhead. A better approach to defend against our proposed attack could be to embed a lightweight integrity check for the \texttt{$if\_id\_pipe\_o$} pipeline register to ensure that instructions meet expected formats, reducing the risk of propagating corrupted instructions through the pipeline~\cite{hamming1950error,krstic2016enhanced}. Likewise, incorporating handshaking protocols to enhance stability and prevent premature latching on the critical paths identified in this study may ensure consistent state transitions~\cite{kommerling1999design}. However, evaluating such defenses lies beyond the scope of this work.

\vspace{-0.35em}
\subsection{Profiling Pre- and Post-Silicon Discrepancies}
\vspace{-0.25em}
To address the deviations between pre- and post-silicon results, a possible solution could be to conduct small-scale device profiling experiments that adjust the glitch offset and width in both environments, measuring the resulting shifts between pre- and post-silicon. This insight could also guide the refinement of pre-silicon simulations based on post-silicon capabilities. However, exploring an automated way to measure deviations between pre- and post-silicon results (by incorporating hardware-specific factors such as clock skew, signal integrity, and FPGA delays into the simulations) is presently beyond the scope of this work.
\vspace{-0.35em}
\subsection{Device Migration Considerations}
\vspace{-0.25em}
We tested CRAFT on a specific RISC-V processor; however, our findings and methodology are adaptable to other processors within the RISC-V family due to their common base architecture. Expanding the analysis to processors with different pipeline configurations offers the chance to tackle new challenges, such as managing variations in pipeline stages, handling data and control hazards, and adjusting to different instruction fetch and execution mechanisms. Additionally, transitioning to processors with deeper pipelines or out-of-order execution units presents an opportunity to refine the pre-silicon characterization approach for more effective handling of unique processing flows.

When migrating to a new hardware platform, such as switching FPGA devices or upgrading processor architectures, it is essential to re-run fault simulations to address design variations. These variations—such as differences in the netlist, logic delays, and timing characteristics—can affect fault injection results \textit{e.g.,} the reported glitch offset and width parameters from one device may not map exactly to a new device due to differences in clock skew or signal integrity. Repeating fault simulations on the new device ensures that previously identified vulnerabilities remain valid and allows for necessary adjustments to mitigate any new risks introduced by the migration. This is an area for future work.
\section{Conclusion}


This work uncovered critical vulnerabilities in RISC-V softcore processors under clock glitch attack, revealing vulnerabilities such as instruction skips, data corruption, and illegal control flow execution.  Our pre-silicon fault analysis of the cv$32$e$40$x pipeline ranks instructions based on their susceptibility at various pipeline stages. Our findings highlight pivotal attack vectors and emphasize the importance of pre- and post-silicon validation, stricter timing constraints, and improved exception handling. Future research could extend these techniques to other architectures, with an emphasis on developing automated, dynamic countermeasures.

\vspace{-0.5em}
\begin{acks}
This work is supported by the Office of Naval Research (ONR) grant N00014-23-1-2103. The views, opinions and/or findings expressed are those of the authors and should not be interpreted as representing the official views or policies of the Department of Defense or the U.S. Government. We used GPT-4.0 to assist with editorial improvements.
\end{acks}

\bibliographystyle{ACM-Reference-Format}
\bibliography{References.bib}

\end{document}